\documentstyle[preprint,aps]{revtex}

\begin{document}

\draft

\title{Generalization of the Maier-Saupe theory of the nematics within
Tsallis Thermostatistics}

\author{O. Kayacan, F. B\"{u}y\"{u}kk{\i}l{\i}\c{c} and D. Demirhan}
\address{Department of Physics, Faculty of Science, Ege University, 
35100 Bornova Izmir, Turkey}

\maketitle

\begin{abstract}
In this study, one of the mean-field theories in
nematics, the Maier-Saupe theory (MST), is generalized within Tsallis
Thermostatistics (TT). The variation of the order parameter versus
temperature has been investigated and compared with experimental data for
PAA (p-azoxyanisole). So far we believe that this is the first attempt of
the application \ of TT in liquid crystals. It is well known that MST fails
to explain the experimental data for some of the nematics, one of which is PAA.
However generalized MST (GMST) is able to account for the experimental data
for a long range of temperatures. Also in this study, the effect of
nonextensivity is shown for various values of the entropic index.

\noindent
{\it PACS Number(s): 05.20.-y, 05.70.-a, 61.30.-v, 61.30.Cz, 61.30.Gd}
\end{abstract}

\newpage

\section{introduction}

There is a growing interest in the nonextensive generalization of the conventional
Boltzmann -Gibbs statistical mechanics. Tsallis Thermostatistics (TT) is one
of these generalization attempts [1] and in this study, we generalize the
Maier-Saupe theory (MST) [2], a mean field theory especially used in nematic
liquid crystals, and the variation of the order parameter of PAA
(p-azoxyanisole, a nematic liquid crystals) versus temperature is
investigated within TT formalism. MST fails to explain the experimental data
for some of the nematics and PAA is one of these nematics. It is well known
that Boltzmann-Gibbs statistics (BG) seems to fail for the systems, (i) with long range microscopic interactions,
(ii) with long range memory effects and (iii) which evolves in a multifractal space-time.

Therefore there is a tendency for nonextensive formalisms. It is seen in the
literature that from 1988 up to present days, TT not only has been applied
to various concepts of thermostatistics [3-20], but also achieved to solve
some physical systems where BG statistics is known to be inadequate. Amongst
them, stellar polytropes [21], Levy-like anomalous diffusions [22],
two-dimensional Euler turbulence [23], solar neutrino problem [24], 
velocity distributions of galaxy clusters [25], fully developed turbulence [26], 
electron-positron and other high-energy collisions [27] 
and anomalous diffusion of Hydra viridissima [28] could be mentioned. 

The axioms of TT are the following
1) the entropy of the physical system is defined by 
\[
S_{q}=-k\frac{1-\sum_{i=1}^{W}p_{i}^{q}}{1-q} 
\]

where k is a constant, p$_{i}$ is the probability of the system in the i th
microstate, W is the total number of configurations and q is the entropic
index (q$\in R$) which is a measure of the nonextensivity of the system. It
is important to note that if q$\rightarrow 1$,\ then S$_{1}=-k_{B}%
\sum_{i}p_{i}\log p_{i}$, which is the well known Shannon entropy.

\noindent
2) q-expectation value of an observable O is given by 
\[
<O>_{q}=\sum_{i=1}^{W}p_{i}^{q}O_{i}
\]

which again recovers the conventional expectation value
when q=1.

\section{The Maier-Saupe theory}

An approach that has proved to be useful in developing a theory of long
range orientational order and the                                                                                                                                                                                                                                                                                                                                                                                                                                                                                                                                \qquad \qquad

where V is the molar volume, A is a constant independent of pressure, volume
and temperature, \ $\theta _{i}$\ is the angle between the long molecular
axis and the preferred axis, s is the order parameter and given by 
\begin{equation}
s=\frac{1}{2}<3\cos ^{2}\theta _{i}-1>.
\end{equation}

The internal energy per mole is 
\begin{equation}
U=\frac{N}{2}\frac{\int_{0}^{1}u_{i}\exp (-u_{i}/kT)d(\cos \theta _{i})}{%
\int_{0}^{1}\exp (-u_{i}/kT)d(\cos \theta _{i})}=-\frac{1}{2}NkTBs^{2}
\end{equation}

where B=A/(kTV$^{2}$), (A=13$\times 10^{-9}$ \ erg.cm$^{6}$, V=225 cm$^{3}$\
forPAA), N is the Avogadro number, T is temperature.

The partition function for a single molecule is given by 
\begin{equation}
z_{i}=\int_{0}^{1}\exp (-u_{i}/kT)d(\cos \theta _{i})
\end{equation}

so that the entropy is 
\begin{equation}
S=-Nk\left[ \frac{1}{2}Bs(2s+1)-\log \int_{0}^{1}\exp (\frac{3}{2}Bs\cos
^{2}\theta _{i})d(\cos \theta _{i})\right] .
\end{equation}

Then the Helmholtz free energy is 
\begin{eqnarray}
F &=&U-TS  \nonumber \\
&=&NKT\left[ \frac{1}{2}Bs(s+1)-\log \int_{0}^{1}\exp (\frac{3}{2}Bs\cos
^{2}\theta _{i})d(\cos \theta _{i})\right] .
\end{eqnarray}

The condition for equilibrium is given by 
\[
\left( \frac{\partial F}{\partial s}\right) _{V,T}=0 
\]

or 
\begin{equation}
3s\frac{\partial }{\partial s}<\cos ^{2}\theta _{i}>-3<\cos ^{2}\theta
_{i}>+1=0.
\end{equation}

This equation is satisfied when 
\begin{equation}
<\cos ^{2}\theta _{i}>=<\cos ^{2}\theta >=\frac{2s+1}{3}.
\end{equation}

The theoretical curve for s versus T could be determined from (8) namely the
equilibrium condition.

\section{Generalization of the Maier-Saupe theory}

In the frame of TT, the generalized partition function is written as 
\begin{equation}
z_{q}=\int_{0}^{1}\left[ 1-(1-q)\beta u_{i}\right] ^{\frac{1}{1-q}}
\end{equation}

where $\beta =1/kT$. The average value of cos$^{2}\theta _{i}$ is 
\begin{equation}
\overline{\cos ^{2}\theta _{i}}=\frac{\int_{0}^{1}\cos ^{2}\theta _{i}\exp
(-u_{i}/kT)d(\cos \theta _{i})}{\int_{0}^{1}\exp (-u_{i}/kT)d(\cos \theta
_{i})}.
\end{equation}

For the sake of simplicity, we replace cos$\theta _{i}$ with x$_{i}$. In
Eq.(10), let us write 
\begin{eqnarray}
\exp _{q}(-u_{i}/kT) &=&\left[ 1-(1-q)\beta u_{i}\right] ^{\frac{1}{1-q}} 
\nonumber \\
&=&\left[ 1+(1-q)\frac{Bs}{2}(3x_{i}^{2}-1)\right] ^{\frac{1}{1-q}}.
\end{eqnarray}

Substituting Eq.(11) into Eq.(10), we obtain 
\begin{equation}
\overline{x^{2}}=\overline{x_{i}^{2}}=\frac{\int_{0}^{1}x_{i}^{2}\left[
1+(1-q)\frac{Bs}{2}(3x_{i}^{2}-1)\right] ^{\frac{1}{1-q}}dx_{i}}{\int_{0}^{1}%
\left[ 1+(1-q)\frac{Bs}{2}(3x_{i}^{2}-1)\right] ^{\frac{1}{1-q}}dx_{i}}.
\end{equation}

To integrate the terms in Eq.(12), the series expansions of exp$%
_{q}(-u_{i}/kT)$ is written\ in terms of q and then the leading two terms
are taken: 
\begin{eqnarray}
\exp _{q}(-u_{i}/kT) &\cong &\exp \left[ \frac{1}{2}Bs(3x_{i}^{2}-1)\right] 
\nonumber \\
&&-\frac{1}{8}\exp \left[ \frac{1}{2}Bs(3x_{i}^{2}-1)\right]
B^{2}s^{2}(3x_{i}^{2}-1)^{2}(1-q).
\end{eqnarray}
Then, we have solved equation (12) and plotted s=f(T) curve for various
values of q. In Figure 1, the order parameter versus temperature for various
values of q are presented. As seen from Fig.1, GMST is in good aggreement
with the experimental data for q=0.988. It seems that with a small departure
from the standard theory one could be able to explain some of the
experimental data available in the study of nematics. GMST recovers MST when
q=1. Near the nematic-isotropic phase transition temperature, the aggreement
between GMST and the experimental data is broken. In Table 1, the variations
of the density of PAA with respect to temperature are given. In addition, in
the recent studies, the entropic index is related to the number of particles
of the system [29,30]. Then a better aggreement with the experimental data
could be obtained, which will be the subject of a forthcoming study.

From GMST, the value of the order parameter \ s at T$_{NI}$ is determined as
0.4 which is in good aggreement with the experimental values [31-35],
whereas 0.443 is given for s in MST.

\section{Conclusion}

\qquad MST is the most widely used molecular field theory in the nematic
liquid crystals. Although this theory is successful in the study of some of
the nematics, it fails for some others, one of which is PAA, p-azoxyanisole.
In this study, it is observed that, after the generalization of MST, the
variation of the order parameter versus the temperature shows a better
aggremeent with the experimental data. Fig.1 exhibits the effect of
nonextensivity. It is remarkable to note that the value of s at T$_{NI}$\ is
0.4 which is given in refs.[31,32,33].

Another important point is the variation of the density of PAA with respect
to the temperature[2], presented in Table 1. As temperature increases, the
density of PAA decreases. Moreover, as mentioned earlier, in recent studies,
the entropic index is related to the number of particles of the system
[29,30]. Therefore it is expected that q, the entropic index, could be
related to the density, which will be the subject of another study in near
future and the variation of s near T$_{NI}$, nematic-isotropic phase
transition temperature, could be explained hopefully by GMST.

Following this study, a tendency to investigate liquid crystals within
nonextensive formalisms could be expected.

\section*{Acknowledgements}

One of us (O.K.) would like to thank U. Tirnakli for valuable comments and
suggestions.

\newpage

\section*{REFERENCES}

\bigskip

[1] C. Tsallis, J. Stat Phys. 52, 479 (1988); E.M.F. Curado, C. Tsallis, J.
Phys. A24, L69 (1991); corrigenda : 24, 3187 (1991); 25, 1019 (1992).

[2] Maier, W. and Saupe, A., A15, 287 (1960).

[3] N. Ito, C. Tsallis, Nuovo Cimento D11, 907 (1989).

[4] R.F.S. Andrade, Physica A175, 285 (1991); A203, 486 (1994); 
U. Tirnakli , D. Demirhan, F. B\"{u}y\"{u}kk\i l\i \c{c}, Acta. Phys. Pol. A91,
1035 (1997).

[5] A.M. Mariz, Phys. Lett. A165, 409 (1992); J.D. Ramshaw, Phys. Lett.
A175, 169 and 171 (1993).

[6] F. B\"{u}y\"{u}kk\i l\i \c{c}, D. Demirhan, Phys. Lett. A181, 24 (1993);
F.B\"{u}y\"{u}kk\i l\i \c{c}, D.Demirhan, A. G\"{u}le\c{c}, Phys. Lett.
A197, 209 (1995).

[7] A. Plastino, A.R. Plastino, Phys. Lett. A177, 177 (1993).

[8] A.R. Plastino, A. Plastino, Physica A202, 438 (1994).

[9] E.P. da Silva, C. Tsallis, E.M.F. Curado, Physica A199, 137 (1993).

[10] A. Plastino, C. Tsallis, J. Phys. A26, L893 (1993).

[11] D.A. Stariolo, Phys. Lett. A185, 262 (1994).

[12] A.R. Plastino, A. Plastino, C. Tsallis, J. Phys. A27, 5707 (1994).

[13] F. B\"{u}y\"{u}kk\i l\i \c{c}, D. Demirhan, Z. Phys. B99, 137 (1995).

[14] S. Curilef, C. Tsallis, Physica A215, 542 (1995).

[15] F. B\"{u}y\"{u}kk\i l\i \c{c}, D. Demirhan, U. Tirnakli , Physica
A238, 285 (1997).

[16] C. Tsallis, F.C. Sa Barreto, E.D. Loh, Phys. Rev. E52, 1447 (1995).

[17] U. Tirnakli , F. Buyukk\i l\i \c{c}, D. Demirhan, Physica A240, 657
(1997).

[18] A.K. Rajagopal, Phys. Rev. Lett. 74, 1048 (1995); Physica B212, 309
(1995).

[19] A.K. Rajagopal, Phys. Rev. Lett. 76, 3469 (1996).

[20] U. Tirnakli , S.F. \"{O}zeren, F. B\"{u}y\"{u}kk\i l\i \c{c}, D.
Demirhan, Z. Phys. B104, 341 (1997).

[21] A.R. Plastino, A. Plastino, Phys. Lett. A174, 384 (1993).

[22] P.A. Alemany, D.H. Zanette, Phys. Rev. E49, R956 (1994); C. Tsallis,
S.V.F. Levy, A.M.C. Souza, R. Maynard, Phys. Rev. Lett. 75, 3589 (1995);
D.H. Zanette , P.A. Alemany, Phys. Rev. Lett. 75, 366 (1995); M.O. Careres,
C.E. Budde, Phys. Rev. Lett. 77, 2589 (1996); D.H. Zanette, P.A. Alemany,
Phys. Rev. Lett. 77, 2590 (1996).

[23] B.M. Boghosian, Phys. Rev. E53, 4754 (1996).

[24] G. Kaniadakis, A. Lavagno, P. Quarati, Phys. Lett. B369, 308 (1996).

[25] A. Lavagno, G. Kaniadakis, M.R. Monteiro, P. Quarati, C. Tsallis,
Astrophys. Lett. and Comm. 35, 449 (1998).

[26] T. Arimitsu and N. Arimitsu, Phys. Rev. E 61, 3237 
(2000); T. Arimitsu and N. Arimitsu, J. Phys. A 33, L235 (2000); 
C. Beck, Physica A 277, 115 (2000); C. Beck, G.S. Lewis and H.L. Swinney, 
Phys. Rev. E 63, 035303 (2001).

[27] D.B. Walton and J. Rafelski, Phys. Rev. Lett. 84, 31 
(2000); G. Wilk and Z. Wlodarcsyk, Nucl. Phys. B  (Proc. Suppl.) 75A, 191 
(1999); G. Wilk and Z. Wlodarczyk, Phys. Rev. Lett. 84, 2770 (2000); 
M.L.D. Ion and D.B. Ion, Phys. Lett. B 482, 57 (2000); 
I. Bediaga, E.M.F. Curado and J. Miranda, Physica A 286, 156 (2000); 
C. Beck, Physica A 286, 164 (2000).

[28] A. Upadhyaya, J.-P. Rieu, J.A. Glazier and Y. Sawada, 
Physica A 293, 549 (2001).

[29] S. Abe, S. Martinez, F. Pennini, A. Plastino, Phys. Lett. A278,
249 (2001).

[30] C. Tsallis, R.S. Mendes, A.R. Plastino, Physica A261, 534 (1998).

[31] V. Zwetkoff, Acta Physicochim. URSS 16, 132 (1942).

[32] W. Maier, A. Saupe, Z. Naturforschg. 13a, 564 (1958).

[33] W. Maier, G. Englert, Z. Electrochem. (1960).

[34] P. Chatelain, Bull. Soc. Fran\c{c}. Min. Crist. 78, 262 (1955).

[35] H. Lippmann, Ann. Phys., Lpz. (6) 20, 265 (1957).

[36] Chandrasekhar S., Madhusudana N.V., Journal Phys. Radium 30, c4-24
(1969).

[37] P. Chatelain and M. Germain, C.R.hebd. Seances Acad. Sci. 259, 127
(1964).

\newpage

\section*{FIGURE AND TABLE CAPTIONS}

\noindent
\textbf{Table 1:}~ The variation of the density of PAA versus
temperature [2].\\

\noindent
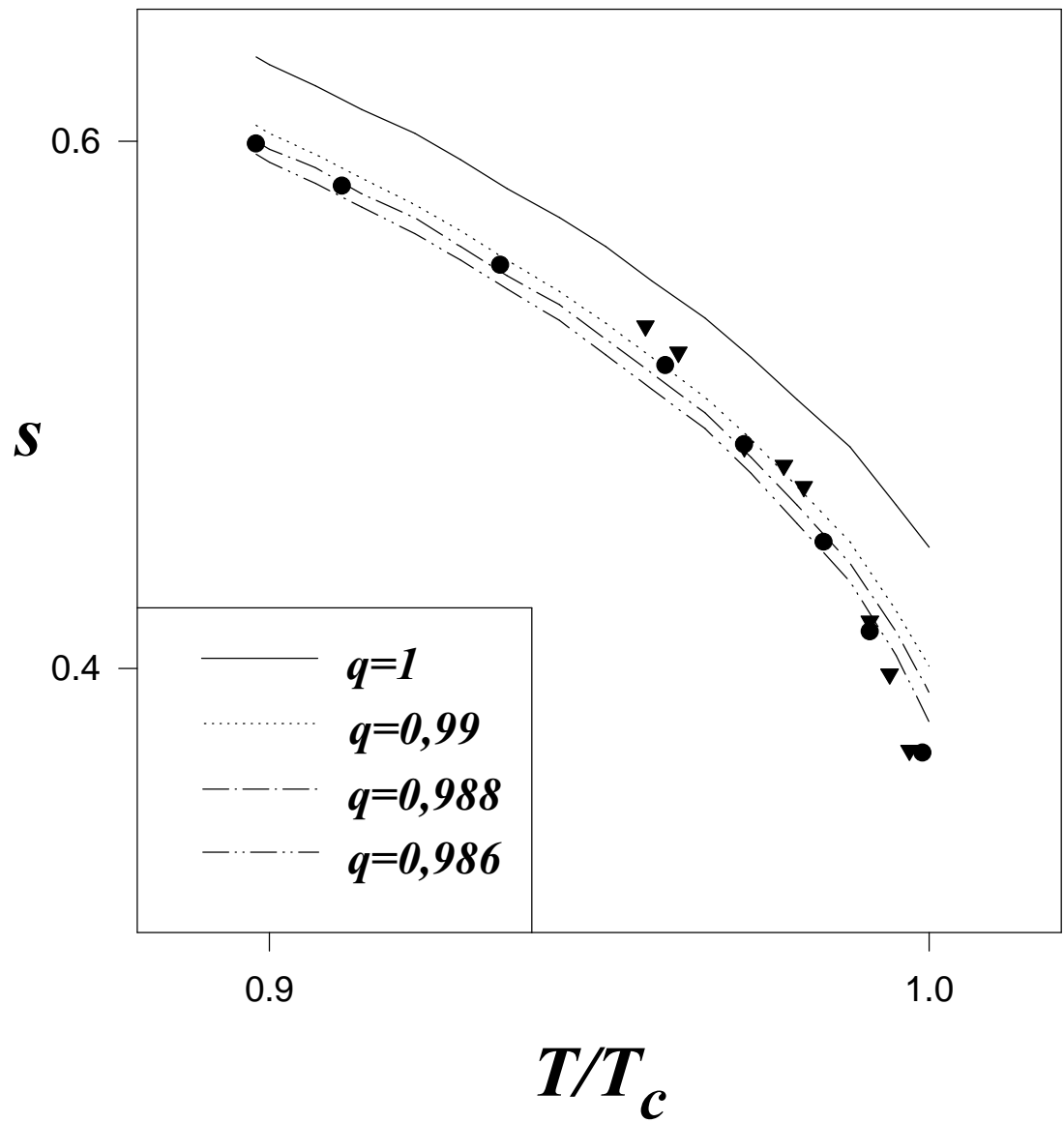
\textbf{Figure 1:}~ Plot of the order parameter versus reduced
temperature T/T$_{c}.$ Continuous line:Maier-Saupe approximations (q$%
\rightarrow 1$). Dashed lines:GMST approximation for various values of q.
Filled circles:optical data [36]. Filled triangles: from refractive index
measurements [37].

\newpage

\section*{TABLE 1.}

\begin{center}
\begin{tabular}{|c|c|c|c|}
\hline
T($^{\circ }$C) & $\rho (g.cm^{-3})$ & T($^{\circ }$C) & $\rho (g.cm^{-3})$
\\ \hline
100 & 1.1824 & 134 & 1.1504 \\ \hline
105 & 1.1781 & 135 & 1.1490 \\ \hline
110 & 1.1737 & 135.5 & 1.1482 \\ \hline
115 & 1.1694 & 135.9 & 1.14392 \\ \hline
120 & 1.1649 & 136 & 1.14380 \\ \hline
122 & 1.1630 & 136.5 & 1.14328 \\ \hline
124 & 1.1611 & 137 & 1.14279 \\ \hline
126 & 1.1592 & 138 & 1.14186 \\ \hline
128 & 1.1572 & 140 & 1.14004 \\ \hline
130 & 1.1551 & 142 & 1.13830 \\ \hline
132 & 1.1528 & 144 & 1.13659 \\ \hline
133 & 1.1516 &  &  \\ \hline
\end{tabular}

\end{center}

\end{document}